# Research progress of artificial intelligence empowered quantum communication and quantum sensing systems[*]

XU Jiaxin[1, 2], XU Lechen[1, 2], LIU Jingyang[1, 2], DING Huajian[1, 2], WANG Qin[1, 2, *]

1. Institute of Quantum Information and Technology, Nanjing University of Posts and Telecommunications, Nanjing 210003, China
2. Key Laboratory of Broadband Wireless Communication and Sensor Network of Ministry of Education, Nanjing University of Posts and Telecommunications, Nanjing 210003, China

## Abstract

Quantum communication and quantum sensing, which leverage the unique characteristics of quantum systems, enable information-theoretically secure communication and high-precision measurement of physical quantities. They have attracted significant attention in recent research. However, they both face numerous challenges on the path to practical application. For instance, device imperfections may lead to security vulnerability, and environmental noise may significantly reduce measurement accuracy. Traditional solutions often involve high computational complexity, long processing time, and substantial hardware resource requirements, posing major obstacles to the large-scale deployment of quantum communication and quantum sensing networks. Artificial intelligence (AI), as a major technological advancement in current scientific landscape, offers powerful data processing and analytical capabilities, providing new ideas and methods for optimizing and enhancing quantum communication and sensing systems. Significant progresses have been made in applying AI to quantum communication and sensing, thus injecting new vitality into these cutting-edge technologies. In quantum communication, AI techniques have greatly improved the performance and security of quantum key distribution, quantum memory, and quantum networks through parameter optimization, real-time feedback control, and attack detection. In quantum sensing, quantum sensing technology enables ultra-high sensitivity detection of physical quantities such as time and magnetic fields. The introduction of AI has opened up new avenues for achieving high-precision and high-sensitivity quantum measurements. With AI, sensor performance is optimized, and measurement accuracy is further enhanced through data analysis. This paper also analyzes the current challenges in using AI to empower quantum communication and sensing systems, such as implementing efficient algorithm deployment and system feedback control under limited computational resources, and addressing complex task environments, dynamically changing scenarios, and multi-task coordination requirements. Finally, this paper discusses and envisions future development prospects in this field.

# 1. Introduction

With the rapid development of information technology, quantum communication, as a new communication technology, has gradually become a research hotspot in the field of communication because of its advantages in security, efficiency and so on. Quantum communication uses the basic principles of quantum mechanics, such as quantum entanglement and no-cloning theorem, to achieve secure transmission of information. However, the practical application of quantum communication systems faces many challenges, such as the defects of system equipment, the resource allocation of large-scale networks and so on. At the same time, as an important branch of quantum information science, quantum sensor technology also shows great potential in the field of high-precision measurement, but in practical applications, it is also restricted by the complexity of system calibration, noise interference and limited measurement accuracy. Traditional solutions are usually computationally complex, time-consuming, and require a large amount of hardware resources, which poses a great challenge to the deployment of large-scale quantum communication and quantum sensor networks.

As an important direction in the field of science and technology, artificial intelligence offers powerful data processing and analysis capabilities, providing new ideas and methods for optimizing and improving quantum communication and quantum sensing systems. For example, in the field of quantum communication, artificial intelligence can quickly and accurately predict the optimal parameters in the quantum communication system, efficiently calibrate the system in real time, and detect possible equipment defects and attacks in the system in real time. In terms of quantum sensing, artificial intelligence can provide an efficient, adaptive and resource-efficient solution for the calibration of large-scale quantum sensors, significantly improving the measurement accuracy. Therefore, the AI-enabled quantum communication and quantum sensing system not only provides the possibility for the breakthrough of quantum communication and quantum sensing technology, but also opens up a new direction for the development of future communication and sensing technology.

## 2. Foundations of Artificial Intelligence

As an important branch of computer science, artificial intelligence (AI) is dedicated to developing theories, methods, technologies and application systems that can simulate, extend and expand human intelligence. Its core is to enable computer systems to learn, reason and make decisions autonomously from massive data without explicit program instructions through a data-driven approach. As one of the core technologies of artificial intelligence, machine learning mainly studies how to enable computer systems to automatically learn rules from data through algorithms, and use these rules to predict or make decisions. It is worth noting that machine learning algorithms are adaptive and can continuously optimize their performance with the input of new data, thus achieving continuous improvement in prediction accuracy.

At the application level, artificial intelligence and machine learning technology have shown great application value in natural science, engineering science, life science, financial economy and other fields. For example, in the physical sciences[1,2], artificial intelligence and machine learning provide new methodological support for the analysis of complex experimental data and the discovery of physical laws.[3] Artificial intelligence algorithms have been widely used in quantum system optimization, quantum state recognition, quantum error correction and other key tasks in the field of quantum science and technology, significantly improving the reliability and security of quantum information processing (see Fig. 1).

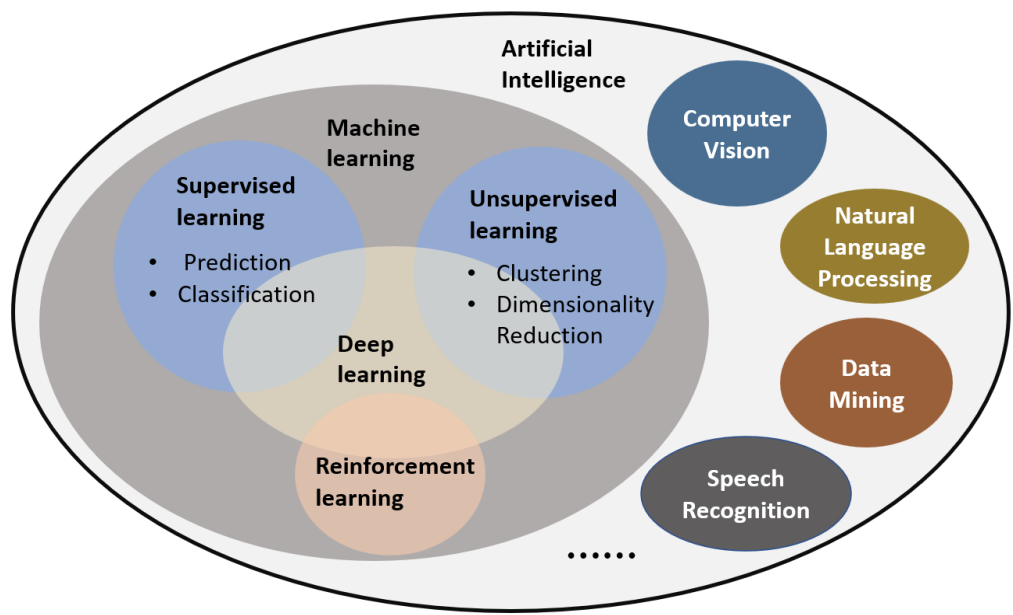

**Figure 1.** An overview of artificial intelligence.

From a methodological point of view, machine learning algorithms can be divided into three categories according to their learning paradigms: supervised learning, unsupervised learning, and reinforcement learning. Supervised learning is suitable for classification and regression tasks by training the model with labeled data sets and establishing the mapping relationship between input and output. Unsupervised learning is dedicated to discovering the internal structure and distribution characteristics of unlabeled data, and has unique advantages in clustering analysis and dimensionality reduction. In addition, reinforcement learning (RL)[4], as an emerging machine learning method, is inspired by behavioral psychology and neuroscience. Through continuous interaction with the environment, the agent optimizes the decision-making strategy or learns the value function, and maximizes the cumulative reward with the help of trial and error and feedback mechanisms, so as to gradually master the ability to make optimal decisions in complex environments (see Fig. 2).

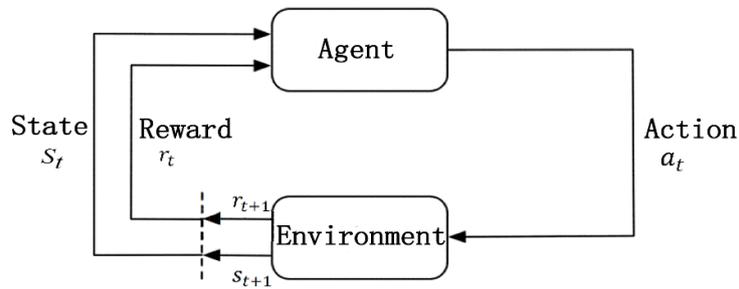

**Figure 2.** Schematic diagram of reinforcement learning.

As an important branch of machine learning, deep learning (DL) realizes high-level abstract representation of data by constructing neural network models with multi-layer nonlinear transformations. This multi-level feature learning mechanism makes deep learning show significant advantages in dealing with high-dimensional and nonlinear data, and has achieved remarkable results in image recognition, speech processing, natural language processing and other fields[5,6]. The neural network (NN)[7] is a complex nonlinear model that simulates the working mode of human brain neurons, which is the basis of deep learning. A neural network consists of a hierarchy of neurons (also called nodes or units), each of which is connected to neurons in the preceding and following layers, processes inputs through weights and activation functions, and produces outputs. As a special recurrent neural network architecture, long short-term memory (LSTM)[8] introduces cell state and three gating mechanisms (forget gate, input gate and output gate) to better control the flow of information, effectively solves the long-term dependency problem in sequential data modeling, and shows excellent performance in time series prediction and natural language generation.

From the perspective of algorithm system, artificial intelligence and machine learning techniques cover a wealth of algorithm models, including but not limited to regression algorithm[9], decision tree algorithm[10], Bayesian algorithm[11], clustering algorithm[12], dimensionality reduction algorithm[13] and ensemble algorithm[14]. Through organic combination and collaborative optimization, these algorithm models provide strong technical support for intelligent decision-making of complex systems.

The rapid development of artificial intelligence and machine learning technology is profoundly changing the paradigm of scientific research and the mode of social production. They not only innovate the traditional data processing and analysis methods, but also provide new ideas and tools for solving complex system problems. In the future, with the innovation of algorithms, the improvement of computing power and the continuous accumulation of data resources, artificial intelligence and machine learning are expected to achieve breakthrough applications in more fields and inject new impetus into the sustainable development of human society.

# 3. Application of Artificial Intelligence in Quantum Communication System

3.1 Quantum key distribution

As one of the core technologies of quantum communication, quantum key distribution (QKD), based on the basic principles of quantum mechanics, can achieve unconditionally secure communication in theory. This technology uses photons as the information carrier, establishes a shared random key string between the two communication parties through the quantum channel, and combines with the one-time pad encryption scheme to construct a communication system with information-theoretic security.

Since Bennett and Brassard[15] proposed the first QKD protocol, namely BB84 protocol in 1984, researchers have proposed E91 protocol device independent QKD[16], BBM92 protocol[17], B92 protocol[18], device independent quantum key distribution (DI-QKD) protocol[19], etc. These protocols vary in security level, device requirements, protocol process, and actual performance, providing many options for QKD implementation in different application scenarios. From the perspective of technical implementation, QKD is mainly divided into discrete-variable QKD (DV-QKD) and continuous-variable QKD (CV-QKD). DV-QKD uses discrete states of qubits (such as the polarization state of a single photon) to encode information. For example, the horizontal polarization state and the vertical polarization state of a single photon can be assigned to binary 0 and 1, respectively. Typical protocols include BB84-QKD protocol, measurement device independent QKD (MDI-QKD) protocol[20], twin-field QKD (TF-QKD) protocol [21], mode-pairing quantum key distribution protocol[22, 23]. CV-QKD uses continuous variables of a quantum system to encode information, for example, to transmit key information by modulating continuous values of the amplitude and phase of a light field. Common protocols include gaussian modulation protocols and discrete modulation protocols.

In practical limited data conditions, the performance of QKD system is highly dependent on the optimization of key parameters, such as the selection probability of X basis or Z basis, the strength of signal state and decoy state, etc. Traditionally, the optimization of these parameters usually relies on search algorithms, which are accurate but computationally complex, time-consuming, and hardware-intensive, posing a significant challenge to real-time QKD systems and large-scale QKD networks. To solve this problem, researchers have proposed a variety of optimization methods based on machine learning, which significantly improve the efficiency and practicability of parameter optimization. Ding et al.[24] used the random forest (RF) algorithm instead of the traditional search algorithm to construct a general model for both MDI-QKD and BB84-QKD protocols, which directly predicts the optimal parameters based on any given system conditions with limited data. Numerical simulations show that the proposed method can achieve an optimal secure key rate of more than 99% compared with the traditional search method, and has a good application prospect in future QKD applications. At the same time, Wang and Lo[25] proposed a method to directly predict the optimal parameters of QKD system using neural networks. Lu et al.[26] proposed a novel back propagation neural network (BPNN), which can not only predict the optimal parameters with less computing resources and faster speed, but also solve the system calibration problem in large-scale MDI-QKD networks. BPNN significantly improves the utility of the system by utilizing partially discarded data generated during communication to enable real-time system calibration without the need for additional equipment or a full system scan. In addition, Dong et al.[27] used the extreme gradient boosting (XGBoost) algorithm to predict the optimized parameters of TF-QKD, and compared the performance with RF and BPNN. The results show that XGBoost is slightly better than RF and BPNN in the efficiency and accuracy of parameter prediction, which provides strong support for the real-time optimization of future quantum key distribution networks. Ren et al. First applied machine learning method to QKD system to realize optimal protocol selection and system parameter optimization[28]. Through comparative analysis, the random forest method shows significant advantages in accuracy, robustness and efficiency, making it a strong candidate for real-time configuration of optimal protocols and system parameters in future large-scale multi-user QKD networks.

In practical applications, QKD systems require efficient real-time feedback control mechanisms to cope with external environmental disturbances and instability of internal components of the system. Although the traditional "scanning-and-transmitting" calibration scheme can provide accurate parameter compensation, the calibration process takes a lot of time, which will greatly reduce the efficiency of key transmission. To this end, Liu et al.[29] proposed a QKD phase modulation scheme based on LSTM. In this scheme, the physical parameters of the device are predicted in advance by the LSTM network, and the phase calibration is performed in real time. In the BB84-QKD system, the scheme has been verified for 48h at 50 km and 150 km transmission distances (see Fig. 3). The experimental results show that the LSTM-based QKD system can maintain the same quantum bit error rate (QBER) level as the traditional scan-transmit scheme while improving the transmission efficiency by at least 33% after 48h of continuous operation, which fully demonstrates the long-term reliability and stability of the machine learning model.

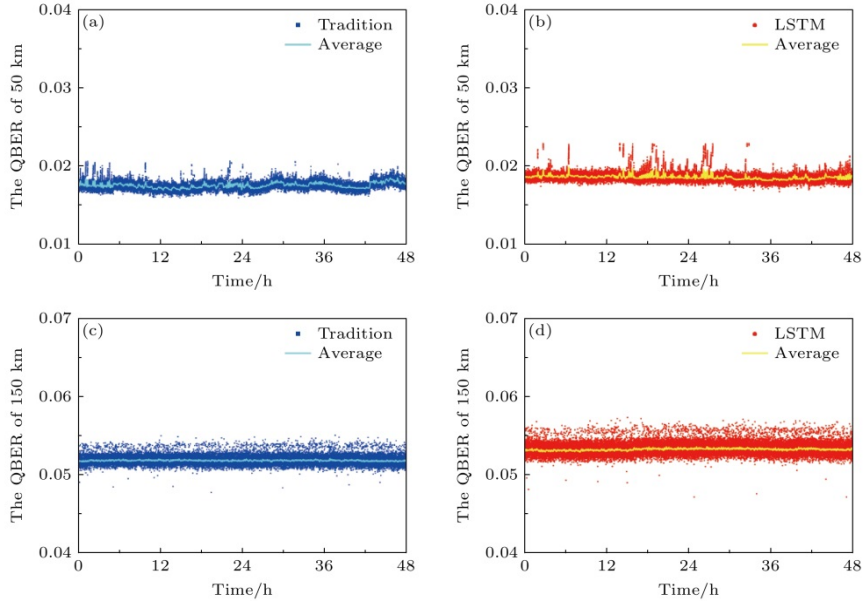

**Figure 3.** Comparisons of QBER between applying traditional scanning-and-transmitting program and using LSTM model for the same QKD system[29].

In order to solve the challenges of measurement equipment or detection, Zhang et al.[30] applied the LSTM model to the MDI-QKD system, successfully predicted the phase drift between two users, and realized real-time active compensation, which significantly improved the reference frame calibration accuracy of the system. TF-QKD protocol has attracted much attention because of its superior secure key rate and transmission distance, but its implementation depends on accurate global phase information. Due to the rapid phase fluctuation in long-distance optical fiber transmission, traditional phase calibration methods (such as time division multiplexing) are inefficient. To address this issue, Liu et al. proposed a neural network-based phase drift prediction technique implemented on a programmable gate array. By loading the double-layer LSTM network into the field-programmable gate array (FPGA), the accurate prediction and active feedback control of the phase drift of the TF-QKD system over a transmission distance of 500 km were realized, and the prediction efficiency was as high as 85%, which greatly improved the transmission efficiency of the TF-QKD system and promoted the implementation of the high-efficiency TF-QKD system.

The realistic security of QKD system is a hot topic in the field of quantum communication. Due to device imperfections, eavesdroppers may exploit the security risk points of various devices to obtain key information. However, existing security analyses often evaluate each security risk point separately, which brings great challenges to the overall security assessment of QKD system. Most of the existing QKD system evaluation schemes require all devices to be tested and calibrated one by one before or after key transmission, which not only consumes a lot of manpower and material resources, but also reduces the transmission efficiency and practicability of QKD. Xu et al.[32] introduced machine learning algorithm into real-time security monitoring of QKD system for the first time, and realized real-time identification of different equipment defects and attacks with an accuracy of up to 98%. This method not only saves time and cost, but also provides an efficient and practical solution for the security evaluation of QKD system without interrupting the key transmission, which lays a solid foundation for the large-scale application of quantum secure communication network in the future. Some typical applications of artificial intelligence in DV-QKD systems are listed in Tab. 1, including parameter optimization, phase calibration, security monitoring, etc., and the main contributions are summarized and compared.

Table 1. Comparison of artificial intelligence applications in DV-QKD.

| Field of application | Method | Main contribution | References |
| --- | --- | --- | --- |
| Parameter optimization | Random forest | Predicting the optimal parameters of MDI-QKD and BB84-QKD protocols | [24] |
| Parameter optimization | Neural network | Direct prediction of optimal parameters of QKD system | [25] |
| Parameter optimization | Extreme gradient boosting | Predict the optimized parameters of TF-QKD with better efficiency and accuracy than RF and BPNN | [27] |
| Parameter Optimization and System Calibration | Back propagation neural network | System optimal parameters are predicted, while the phase of the MDI-QKD system is calibrated by simulation | [26] |
| Protocol selection and parameter optimization | Random forest | For the first time, machine learning methods are applied to QKD system to achieve optimal protocol selection and system parameter optimization | [28] |
| Phase calibration | Long short-term memory network | Predicting the physical parameters of the device and calibrating the phase of the BB84-QKD system in real time | [29] |
| Phase calibration | Long short-term memory network | Predicting the phase drift of two users in MDI-QKD system, active compensation in real time | [30] |
| Phase calibration | Long short-term memory network | Predicting phase drift of TF-QKD system for active feedback control | [31] |
| Device Defect and Attack Detection | Random forest | Real-time detection of device defects and attacks with up to 98% accuracy | [32] |

The above is about the application of artificial intelligence in DV-QKD system. Artificial intelligence methods also have many applications in CV-QKD system, mainly in the optimization of system performance and the improvement of security. In the aspect of system performance optimization, researchers predict and compensate the phase drift by analyzing the fluctuation of system physical parameters, such as the change of local oscillator intensity, the disturbance of channel, the change of ambient temperature, the jitter of devices, etc. In this method, a standard sequence known to both sides is inserted before the effective transmission data, and the phase offset is obtained by comparing the data at the receiving end with the data at the sending end. Then, using machine learning techniques, a prediction model of the time and phase drift values can be established. When the system transmits an effective sequence, the phase drift value is predicted by the model and loaded on the phase modulator at the transmitting end as a phase compensation angle for phase compensation. This method can well analyze the system phase drift value and can be used as a phase compensation algorithm for optimization. In addition, the original data can also be reconstructed directly by the predicted value, and the phase drift value does not need to be calculated in real time, which reduces the system overhead to a certain extent.

In 2017, Liu et al.[33] developed a method (see Fig. 4) through a support vector regression (SVR) model to optimize the performance and practical security of a QKD system. The SVR model is learned to accurately predict the time-varying physical parameters of a signal. Secondly, the predicted time-varying feedback is used to control the QKD system to achieve the best performance and practical security.

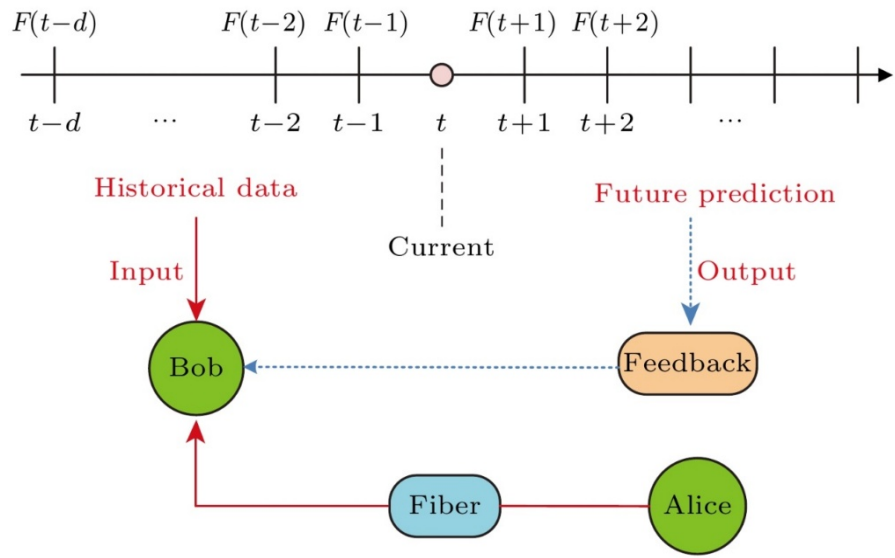

**Figure 4.** Basic idea of SVR to solve the physical parameters prediction problem[33].

In 2019, Su et al.[34] applied the BPNN algorithm to the four-state discrete modulation protocol in CV-QKD, which can adjust the modulation variance to the optimal value, thus ensuring the system security and achieving the best system performance. Numerical results show that the scheme can effectively improve the key rate. In 2020, Liao et al.[35] proposed a new scheme of discrete modulation CV-QKD using machine learning technology, called CV-QKD based on multi-label learning. Specifically, the scheme divides the whole quantum system into a state learning process and a state prediction process. The former is used to train and estimate the classifier, and the latter is used to generate the final key. At the same time, a multi-label classification algorithm is designed as an embedded classifier to distinguish coherent states. The feature extraction of coherent states and the related machine learning indicators of quantum classifiers have been proposed successively. The CV-QKD scheme based on multi-label learning is superior to other existing discrete modulation CV-QKD protocols, such as the four-state protocol and the eight-state protocol, and is also superior to the original Gaussian modulation CV-QKD protocol, and its performance will be further improved with the increase of modulation variance. In 2022, Zhou et al.[36] constructed a neural network to predict the key rate of the discrete modulation CV-QKD protocol, which can quickly predict the secure key rate according to the experimental parameters and results. Compared with the traditional numerical method, the speed of the neural network is improved by several orders of magnitude. Importantly, the predicted key rate is not only highly accurate, but also secure, which allows the secure key rate of discretely modulated CV-QKD to be extracted in real time on a low-power platform, and the scheme can also be extended to quickly calculate the complex secure key rate of various other unstructured quantum key distribution protocols. In the same year, Liu et al.[37] proposed a neural network model combined with Bayesian optimization to predict the secure key rate of discrete modulation CV-QKD protocol in real time. The model is about $10^7$ times faster than the traditional numerical method, and the prediction results have high security and accuracy, which can meet the needs of low power consumption scenarios such as mobile platforms.

In terms of security, CV-QKD systems are threatened by a variety of attack strategies. Existing defense methods usually rely on real-time monitoring modules, but their effectiveness is limited by the accuracy of additional noise estimation and lack of generality. Therefore, in 2020, Mao et al.[38] proposed a quantum attack detection model based on artificial neural network (ANN) (see Fig. 5). By analyzing the feature vector of the attacked pulse, the model realizes the automatic identification and classification of the attack type, and the accuracy and recall are both more than 99%, which significantly improves the security of the system.

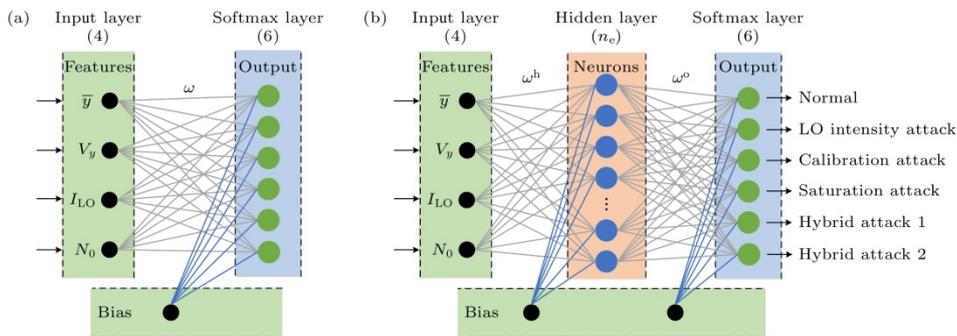

**Figure 5.** ANN-based quantum attack detection model[38]: (a) A linear ANN model without the hidden layer which can only solve linear separable problems; (b) a nonlinear ANN model with a hidden layer to classify different types of quantum attacks.

Subsequently, in 2023, Ding et al.[39] proposed an attack detection scheme based on machine learning, and its implementation process is shown in Fig. 6. Combining the advantages of density-based spatial clustering of applications with noise (DBSCAN) and multiclass support vector machines (MCSVMs), the efficient detection of quantum hacking attacks is realized. The scheme first extracts the feature vectors related to the attack, uses DBSCAN to remove noise and outliers, and then uses the trained MCSVMs to classify and predict whether to generate the final key in real time. Simulation results show that the scheme can effectively detect most attacks, and correct the key rate overestimation problem of CV-QKD system without defense strategy, thus providing a tighter security boundary.

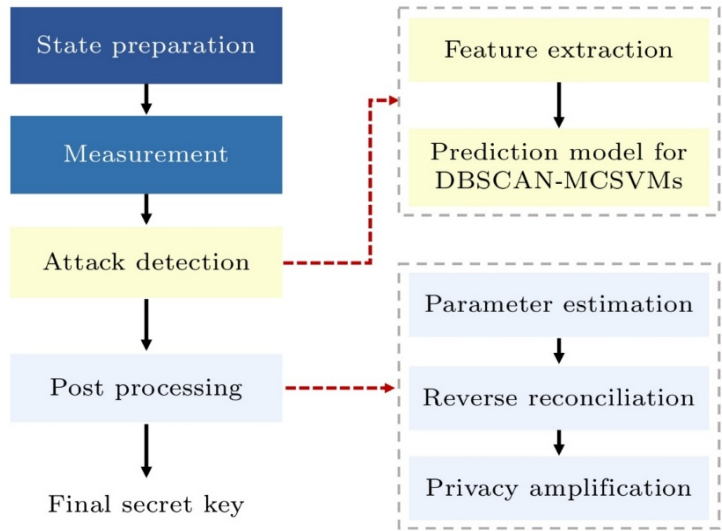

**Figure 6.** Implementation process of a machine-learning-based attack detection scheme[39].

In long-distance CV-QKD experiments, traditional schemes usually rely on the transmission local oscillator, which not only increases the system complexity, but also may introduce security vulnerabilities. Hajomer et al.[40] designed a new long-distance CV-QKD experimental scheme, which successfully realized key distribution over 100 km fibre channel with a total loss of 15.4 dB by locally generating a local oscillator. This breakthrough enables secure key generation against collective attacks in finite-size scenarios by controlling the additional noise caused by phase noise and optimizing the modulation variance through a machine learning framework. This breakthrough not only marks an important milestone for CV-QKD technology under a high-loss budget, but also paves the way for the deployment of large-scale secure QKD networks. In the future, with the continuous development of quantum technology, CV-QKD is expected to be widely used in more high-security and long-distance communication scenarios. Various typical application reports of artificial intelligence in CV-QKD are listed in Tab. 2, and the main contributions are summarized and compared.

Table 2. Comparison of artificial intelligence applications in CV-QKD.

| Field of application | Method | Main contribution | References |
|---|---|---|---|
| Parameter optimization | Support vector regression | Predicting system physical parameters to optimize QKD system performance and security | [33] |
| Parameter optimization | Back propagation neural network | The modulation variance is adjusted to ensure the security of the system and effectively improve the key rate. | [34] |
| Parameter optimization | Machine Learning Framework | Controlling phase noise and optimizing modulation variance for key distribution over 100 km fibre channel | [40] |
| Key Rate Prediction | Multi-label classification algorithm | Discrimination of coherent States by a multi-label classification algorithm, outperforming existing discrete-modulation CV-QKD protocols | [35] |
| Key Rate Prediction | Neural network | Fast prediction of the key rate of discretely modulated CV-QKD protocols with speed and accuracy superior to conventional numerical methods | [36,37] |
| Attack detection | Artificial neural network | Automatic recognition and classification of attack types with over 99% precision and recall | [38] |
| Attack detection | Density clustering and multi-class support vector machine. | Efficient detection of quantum hacking attacks, correction of key rate overestimation, and provision of tighter security bounds | [39] |

3.2 Quantum memory

Quantum memory is the core component of quantum repeater, which is widely used in quantum communication, quantum computation and quantum network, and its performance exceeds the range allowed by passive transmission only. The core of quantum memory is to store a quantum state in a physical system and read it on demand. An ideal quantum memory should have high efficiency, long coherence time and low noise. Common quantum memory mechanisms include electromagnetically induced transparency[41], Raman scheme[42], gradient echo memory[43], and optical cavity reflection[44]. Among them, quantum memory based on solid systems (such as color centers and rare earth ions) has attracted much attention because of its stability, decoherence resistance and scalability[45].

In order to further improve the performance of quantum memory, researchers have introduced machine learning technology into this field, which has significantly improved the optical depth and storage efficiency of the system. Leung et al.[46] optimized the performance of gradient echo memory by combining machine learning with single photon technology through experimental research, and explored its application potential in quantum computing. Due to the complexity of cold atomic ensemble systems, the realization of high optical depth and cryogenic environments often faces great challenges. In this scheme, the atom trapping process is optimized by reinforcement learning, and a high extinction ratio filtering technique is developed to successfully separate single photons from intense pump light. These optimizations significantly improve the optical depth and temperature control accuracy of the cold atom assembly, thereby improving the efficiency and coherence time of the gradient echo memory.

As the basic unit of quantum information processing, the generation and storage technology of single photon is very important for the construction of quantum network. However, there are still many challenges to realize single photon generation and storage with arbitrary wave packet shape. Cai et al.[47] theoretically proposed a general machine learning algorithm with adaptive process (see Fig. 7) to optimize the atom-cavity system and realize high efficiency and high fidelity single photon generation and storage. The algorithm can dynamically adjust the parameters of the quantum system according to the wave function of a single photon, which provides a new technical path for building a flexible and reliable quantum network.

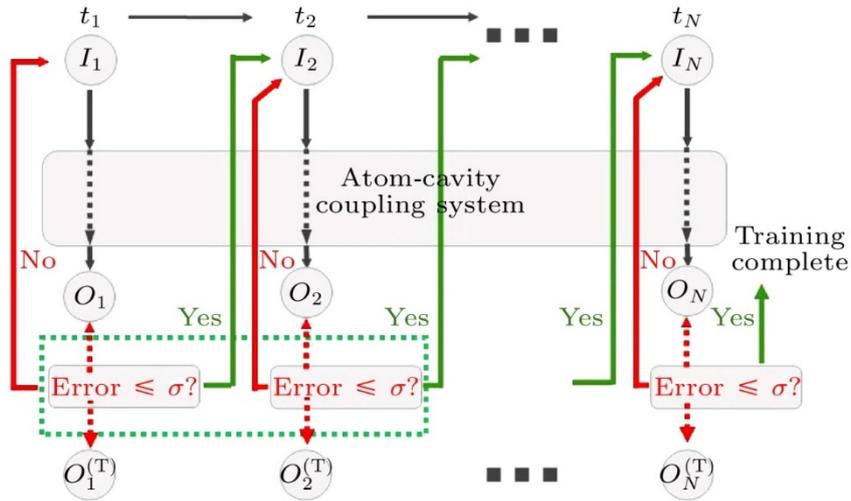

**Figure 7.** Detailed schematic of the ML algorithm[47], the input dataset (discrete control laser pulse) iteratively adjusts itself until the error (feedback in training) between the estimated output and the target value becomes small enough.

Despite the remarkable progress in the research of quantum networks, the coherence time of quantum memories is still a key factor limiting their development. Khatri[48] proposed a design method of optimal entanglement distribution protocol based on the theory of decision process.

Through dynamic programming or reinforcement learning algorithm, the entanglement time on the quantum channel is optimized under the consideration of the current hardware constraints (including the finite coherence time), and the entanglement state between the terminal nodes is ensured to be established before the channel decays. This study provides important theoretical guidance for the physical realization of quantum networks.

In the aspect of quantum memory time optimization, Reiß et al. [49] theoretically used reinforcement learning technology to realize the dynamic optimization of quantum memory time limit (memory cutoff), which provided a flexible strategy for the adjustment of quantum repeater state. A recent study by Robertson et al.[50] has shown that genetic algorithms can be used to optimize the write control of optical memories to handle Gaussian signal pulses. The experimental results show that the pulse energy can be reduced by 30% without sacrificing the efficiency, which significantly improves the storage performance. The high storage efficiency and broadband characteristics of quantum memory are the core elements of future quantum networks. Although solid-state quantum memories have advantages in broadband storage, their storage efficiency is generally low. In response to this challenge, Lei et al.[51] have experimentally demonstrated that the quantum storage efficiency has been improved by nearly six times through the combination of passive optimization and algorithmic optimization, and coherent and single-photon level storage with high signal-to-noise ratio has been achieved. This optimization scheme has wide applicability and can be applied to most solid-state quantum memories to significantly improve storage efficiency while maintaining bandwidth.

3.3 Quantum network

Quantum network plays a vital role in quantum information science, and its applications cover quantum communication, quantum computing, quantum metrology and other fields. One of the central challenges in implementing a quantum network lies in distributing entangled flying qubits (usually implemented in the form of photons) to spatially separated nodes and mapping the entanglement onto stationary qubits (such as matter-based quantum memories) via a quantum interface or transducer. This architecture of separated nodes forms the basis of a quantum network, in which stationary qubits serve as memory units and flying qubits serve as channels for quantum information transmission.

At present, quantum network has been applied to QKD with trusted nodes[52]. Due to the non-reusability of key resources in QKD networks, the allocation of key resources in QKD networks is significantly different from that in traditional networks. Considering the high cost and complexity of QKD network deployment, the multi-tenant model has become an important solution to improve the cost-effectiveness of future QKD networks, especially for organizations with high security requirements. In order to optimize the resource allocation in multi-tenant environment, Cao et al[53] proposed a multi-tenant key distribution algorithm based on reinforcement learning, which can efficiently distribute a variety of network resources including keys. Experimental results show that, compared with the traditional heuristic methods such as random allocation, fitting and best fitting, the proposed algorithm performs well in reducing the blocking probability of tenant requests and improving the utilization of key resources, and reduces the blocking probability by more than half. In addition, Cao et al.[54] further studied the online multi-tenant configuration problem in quantum key distribution networks, proposed three heuristic algorithms (random, adaptive and best-fit online multi-tenant configuration algorithms), and developed a reinforcement learning framework to realize the automatic optimization of the algorithm. The comparison results show that after enough training iterations, the online multi-tenant provisioning algorithm based on reinforcement learning is significantly better than the heuristic method in terms of tenant request blocking probability and key resource utilization. Sharma et al.[55] theoretically proposed a routing and resource allocation scheme based on reinforcement learning to optimize the performance of quantum signal channels in quantum key distribution secure optical networks (see Fig. 8). The core of the scheme is to obtain network state information through a software-defined network controller, and a deep neural network selects a routing and resource allocation strategy according to the state. By maximizing the number of QKD optical path requests, reducing the blocking rate and efficiently utilizing network resources, the proposed scheme significantly improves the network performance in dynamic traffic scenarios compared with traditional methods, and provides a new and effective method for the optimization of QKD optical networks.

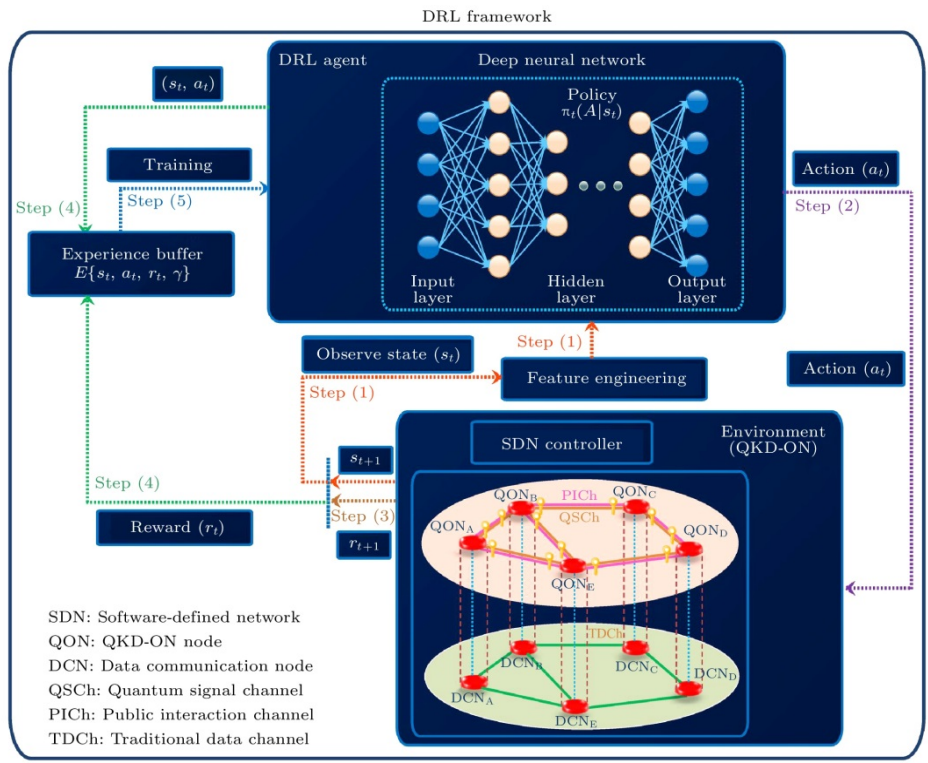

**Figure 8.** An illustration of the proposed deep reinforcement learning framework for the routing and resource assignment in quantum key distribution-secured optical networks[55].

The expansion from point-to-point quantum key distribution to multi-point communication is an inevitable trend for the large-scale development of quantum key distribution networks. Currently, integrating quantum access networks (QANs) into existing ethernet passive optical access networks (EPONs) is a relatively simple and cost-effective implementation. Kang et al.[56] proposed a quantum-secure 10 Gbit/s ethernet passive optical network, and experimentally developed and validated a plug-and-play two-field quantum key distribution architecture that supports up to 64 users with only an untrusted laser and a pair of shared detectors (see Fig. 9). In addition, they proposed a user demand-oriented prediction model based on machine learning to evaluate the key indicators of QAN (such as security key rate, maximum feeder fiber length, etc.). This plug-and-play two-field quantum access network and its machine learning-assisted implementation provide guidance for further experiments and practical deployment of large-scale quantum access networks.

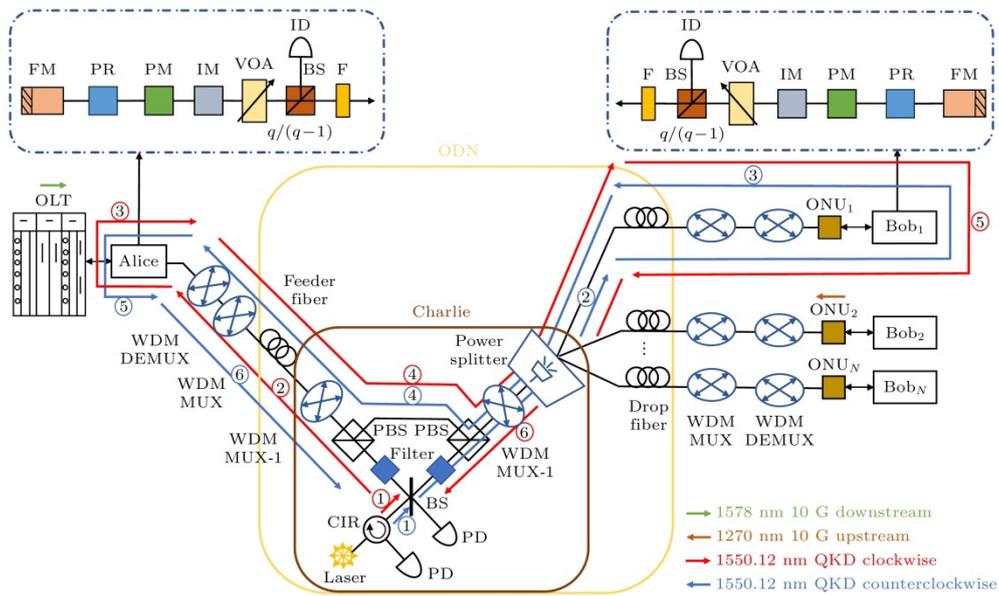

**Figure 9.** Full coexistence architecture of plug-and-play twin-field QAN and 10 G-EPON[56].

At present, the development of quantum networks is at a critical stage of transition from theoretical research to practical application. As the two core technologies of quantum network, quantum memory and trusted relay have their own advantages and application scenarios. The advantage of quantum memory is that it can store quantum information for a long time and read it on demand, thus enhancing the flexibility and robustness of the network. However, its performance still needs to be further improved to meet the needs of large-scale quantum networks. In contrast, trusted relay schemes have made significant progress in long-distance quantum communication, but they rely on the deployment of physical nodes and have limited scalability and flexibility. In addition, due to the need to measure and re-prepare the quantum state at the intermediate node, the security of the trusted relay scheme also faces certain challenges. In the future, the development of quantum networks will rely on the further improvement of quantum memory performance and the optimization of trusted relay schemes to achieve more efficient and secure quantum communication networks.

# 4. Application of Artificial Intelligence in Quantum Sensing System

4.1 Introduction to Quantum Sensing

Quantum sensing is a technology that uses quantum systems, quantum properties or quantum phenomena to measure physical quantities (such as magnetic field, electric field, temperature, pressure, etc.) With high precision. It is based on the basic principles of quantum mechanics, especially the superposition, entanglement and quantum phase evolution of quantum States, so as to achieve measurement accuracy beyond classical sensors. Quantum sensor is a device that uses quantum mechanical properties (such as atomic energy level, photon state or spin of elementary particles) for measurement. The typical representatives of quantum sensors that have entered the practical stage are atomic clock[57], atomic magnetometer[58], atomic interference gravimeter[59], etc. At the same time, the sensor technology based on diamond NV color center[60] and Rydberg atom[61] is gradually becoming an important development direction.

The current mainstream quantum sensing technology can be attributed to the following three types of[62]: 1) the use of quantum objects to measure physical quantities, which are characterized by quantized energy levels (quantum states). Specific examples include electronic, magnetic, nuclear, or vibrational energy levels from superconducting, neutral atoms, trapped ions, or other spin systems. 2) measurement of physical quantities using quantum coherence (i.e., spatial and temporal superposition States with wave properties). 3) the use of quantum resources such as entanglement and squeezing to improve the sensitivity or accuracy of measurement, thus exceeding the statistical limit of classical measurement technology. Most of the early quantum sensing technologies focused on improving the ability to sense small changes in physical quantities. With the realization of quantum entanglement and the development of laser technology, people begin to fully tap the great potential of special quantum states and quantum state manipulation methods in reducing the measurement uncertainty. This kind of technology pays more attention to how to further reduce the measurement uncertainty of the physical quantity to be measured in the case of limited resources. At present, several quantum sensing schemes have been successfully demonstrated in photonic, atomic, nuclear magnetic resonance, and solid-state systems. Compared with the of solid systems and atomic systems[63,64], photons exhibit longer coherence times, lower susceptibility to disturbance, easier controllability and better scalability; therefore photonic systems possess natural advantages for sensing tasks and have gradually attracted more and more attention in recent years.

4.2 Combination of quantum communication system and sensing technology

In recent years, with the rapid development of communication technology, the integration of quantum communication and sensing technology has gradually become a research hotspot. Quantum communication uses quantum states as information carriers to achieve unconditionally secure key distribution in theory. At the same time, sensing technology is also developing. Distributed optical fiber sensing technology uses the light backscattering in the optical fiber or the change of the forward transmission light to monitor the environmental changes, which has the advantages of high spatial resolution and long detection range. The combination of quantum communication system and traditional sensing technology can not only realize the integration of communication and sensing, but also endow the communication network with environmental perception ability without changing the existing quantum communication network architecture, providing new technical means for early warning of earthquakes, landslides and other disasters.

In recent years, many research teams have made important achievements in the field of quantum communication and sensing fusion. Chen et al.[65] realized TF-QKD over 658 km fiber and distributed vibration sensing using the frequency calibration link in the TF-QKD system. This study not only expands the distance record of quantum key distribution, but also successfully locates the vibration source with an accuracy of better than 1 km. Xu et al.[66] proposed an integrated distributed sensing and quantum communication network scheme, which demonstrated the potential of the fusion of quantum communication and sensing by simultaneously implementing continuous-variable quantum key distribution and distributed vibration sensing in optical fibers. Under 10 km standard optical fiber transmission, the key rate of each user is about 0.7 Mbits/s, the vibration response bandwidth is 1 Hz — 2 kHz, and the spatial resolution is 0.2 m. Liu et al.[67] proposed a network architecture integrating downstream quantum access network and optical fiber vibration sensing. By simultaneously encoding the key information of eight users on the sideband of a single laser source, and separating and distributing the key information using a narrow-band filtering network, an average key rate of $1.94 \times 10^4$ bits/s is achieved, and vibration localization with spatial resolutions of 131 m, 25 m, and 4 m at vibration frequencies of 100 Hz, 1 kHz, and 10 kHz is simultaneously achieved.

However, traditional sensing techniques have limitations in accuracy. For example, although distributed optical fiber sensing based on backscattered light can achieve high spatial resolution, its detection accuracy is usually limited by noise due to its dependence on the reflection of weak optical signals, and it is difficult to reach the Heisenberg limit. In addition, traditional optical fiber sensing technologies require high-power light sources (such as erbium-doped fiber amplifiers) for long-distance detection, which not only increases the complexity and cost of the system, but also may cause interference to low optical signals in quantum communication systems. In contrast, quantum sensing technology uses the characteristics of quantum states to provide higher sensitivity than classical measurements, which can significantly improve the sensing accuracy[63,68]. In addition, the use of multi-photon entangled states, single photons, squeezed states and other means of[69–71] can surpass the standard quantum limit, and even theoretically approach the Heisenberg limit.

Distributed quantum sensing is a promising research direction in quantum networks, which can significantly improve the accuracy and efficiency of measurement by performing sensing tasks in a network of multiple measurement nodes. Compared with single-parameter quantum sensing, distributed quantum sensing can measure linear combinations of multiple parameters with high precision, which can be divided into continuous variable scheme[70] and discrete variable scheme[71,72]. However, distributed quantum sensing still faces some challenges. For example, how to measure any unknown parameter with high precision in practical application is a key problem. In the future, with the further development of quantum entanglement technology, quantum network architecture and quantum communication technology, distributed quantum sensing technology is expected to achieve wider applications in quantum communication networks and promote breakthroughs in quantum technology in many fields.

4.3 AI-enabled quantum sensing system

In recent years, artificial intelligence technology has been gradually introduced into the field of quantum sensing, providing new methods and ideas for the calibration, optimization and performance improvement of quantum sensors. Cimini et al.[73] proposed a calibration method of quantum phase sensor based on neural network. This method does not rely on complex theoretical models to describe all the parameters and noise sources of the sensor, but directly calibrates the sensor through neural networks. It is found that the neural network can effectively deal with the uncertainty in the training data, and achieve the measurement accuracy close to the quantum limit in the experiment. In addition, the method is robust to noise, and the performance can be further optimized by adjusting the training parameters. This method provides an efficient, adaptive and resource-efficient solution for the calibration of quantum sensors, especially for the calibration of large-scale quantum devices. In addition, Hentschel and Sanders particle swarm optimization theoretically proposed the application of machine learning to quantum phase estimation, and designed a feedback strategy for interferometer phase estimation through particle swarm optimization (PSO) algorithm, which significantly improved the measurement accuracy and approached the Heisenberg limit. Compared with the traditional BWB (Berry-Wiseman-Breslin) strategy, the strategy generated by PSO shows better performance in both noiseless and noisy conditions.

Reinforcement learning achieves its goal by optimizing the strategy through trial and error through the interaction between the agent and the environment. Xu et al.[75] theoretically proposed the use of reinforcement learning to optimize the control strategy in quantum parameter estimation, and proved the high efficiency and generalization ability of reinforcement learning in this field. By training a neural network to generate control sequences for different parameter values, they avoid the high computational cost of re-optimization every time the parameters are updated. Schuff et al.[76] theoretically optimized the dynamic characteristics of quantum sensors by using the cross-entropy method in reinforcement learning. Based on the quantum chaotic sensor, the measurement accuracy can be further improved and the decoherence can be resisted by optimizing the intensity and position of the nonlinear control pulse. Compared with the traditional periodic control pulse, the optimized control strategy significantly improves the quantum Fisher information (QFI) in both superradiant damping and phase damping decoherence models, and in some cases, the sensitivity is improved by more than an order of magnitude. By visualizing the evolution of quantum States, the study reveals that reinforcement learning uses a strategy similar to spin squeezing, which can adapt to the dynamic characteristics of superradiant damping. Xiao et al.[77] proposed a quantum parameter estimation scheme based on deep reinforcement learning to deal with time-dependent parameter estimation problems. This study geometrically derives the noiseless and noisy boundaries of the QFI for parameter estimation, and designs a reward function associated with these boundaries to accelerate network training and quickly generate quantum control signals (see Fig. 10). Simulation results show that the scheme exhibits good robustness and sample efficiency both in the absence of noise and in the presence of noise (including decoherence and spontaneous emission noise), and reaches the theoretical performance limit. In addition, the study also evaluates the transferability of the scheme when the parameters deviate from the true value, and the results show that the scheme performs well in time-independent parameter estimation, but is more sensitive to noise in time-dependent parameter estimation.

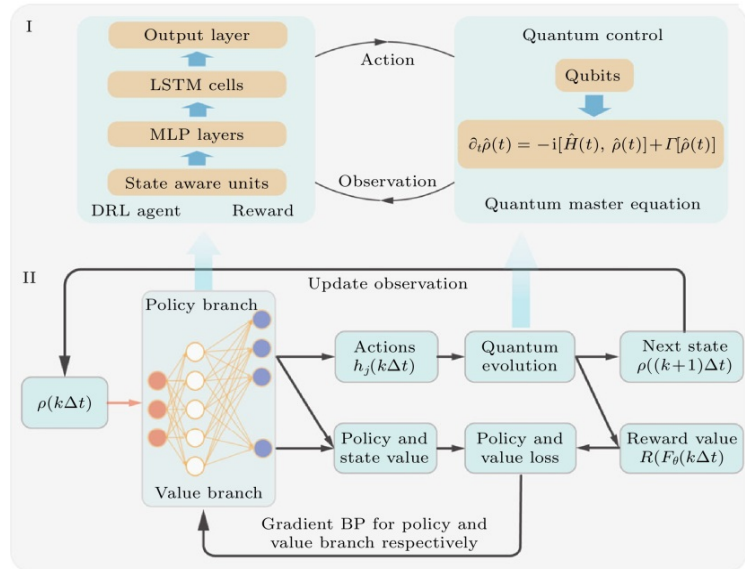

**Figure 10.** Illustration of deep reinforcement learning with (I) agent-environment interaction (II) state-aware policy and value networks with LSTMCells for quantum sensing protocols[77].

Model-aware reinforcement learning (MARL) further integrates the prior knowledge of physical models into the training process, and deals with the non-differential steps in quantum metrology through automatic differentiation techniques, such as measurement and resampling of particle filters. This method not only improves the efficiency of strategy optimization, but also can deal with complex quantum systems. For example, Belliardo et al.[78] theoretically developed a model-based reinforcement learning framework for optimizing Bayesian experiment design in quantum metrology. By combining Bayesian estimation and reinforcement learning, they have achieved high-precision parameter estimation on a variety of quantum platforms, such as NV color centers and photonic circuits. This method not only has advantages in theory, but also demonstrates its efficiency in experimental design through practical application.

Liu et al.[79] used the method of artificial intelligence to experimentally realize the precise detection of multi-frequency microwave based on Rydberg atoms. This work organically combines atomic sensing with deep learning, proposes and realizes the scheme of effectively detecting multi-frequency microwave electric field without solving the master equation, and can achieve high accuracy without too high hardware requirements, which provides an important reference for the cross combination of sensing and neural network, and has important application prospects in communication, radar detection and other fields. Zhou et al.[80] proposed a deep learning based optical quantum sensing scheme to address the challenges of implementing optical quantum sensing in unknown or uncontrolled environments (see Fig. 11). Traditional methods usually rely on the prior knowledge of the target system to achieve the Heisenberg limit of measurement accuracy, but in practical applications, this prior knowledge is often difficult to obtain. Therefore, by combining graph neural network (GNN) and trigonometric interpolation algorithm, the scheme enables optical quantum sensors to achieve Heisenberg limit accuracy in unknown environments through numerical simulation, which provides a new solution for quantum sensing in complex environments. Various applications of artificial intelligence in the field of quantum sensing are summarized and compared in Tab. 3.

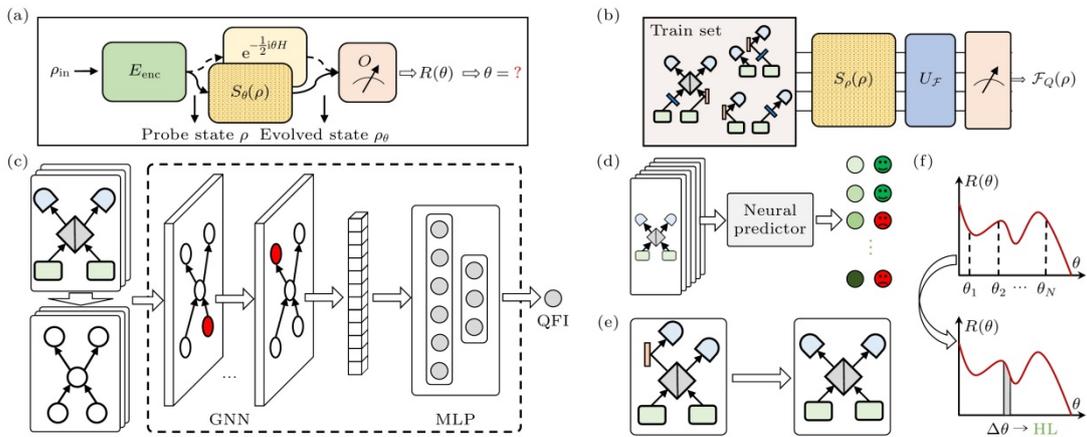

**Figure 11.** Schematic of deep-learning-based quantum sensing[80].

**Table 3.** Comparison of artificial intelligence applications in quantum sensing.

| Field of application | Method | Main contribution | References |
|---|---|---|---|
| Quantum sensor calibration | Neural network | Using neural networks to handle uncertainty in training data to achieve measurement precision near the quantum limit | [73] |
| Parameter estimation | Particle swarm optimization | Automatically design a feedback strategy for interferometer phase estimation with accuracy close to the Heisenberg limit and better than the conventional BWB strategy | [74] |
| Parameter estimation | Reinforcement learning | The neural network is trained to generate control sequences suitable for different parameter values, avoiding the high computational cost of re-optimization at each parameter update | [75] |
| Quantum sensor optimization | Reinforcement learning | Training neural network generation to optimize the dynamics of quantum sensors using cross entropy method of reinforcement learning | [76] |
| Parameter estimation | Deep reinforcement learning | The noiseless and noisy boundaries of the QFI for parameter estimation are derived from a geometric perspective, showing good robustness and sample efficiency under both noiseless and noisy conditions. | [77] |
| Parameter estimation | Model-Aware Reinforcement Learning | Combining Bayesian estimation and reinforcement learning to optimize experimental design in quantum metrology for multiple quantum platforms | [78] |
| Unknown environment | Deep learning | Combining graph neural network and trigonometric interpolation algorithm, the optical quantum sensor achieves Heisenberg limit accuracy in unknown environment. | [80] |
| Microwave detection | Deep learning | This paper presents a scheme for effectively detecting multi-frequency microwave electric field without solving the master equation, which requires low hardware requirements and high accuracy. | [79] |

At present, the application of artificial intelligence algorithm in the field of quantum sensing mainly focuses on single quantum sensing system. With the help of artificial intelligence technologies such as neural networks and reinforcement learning, these systems have achieved efficient sensor calibration, performance optimization, and significant improvement in measurement accuracy. However, for the distributed quantum sensing system, it involves the cooperation between multiple nodes, and more importantly, the scalability and stability of the quantum network constitute the core bottleneck restricting the development of distributed quantum sensing technology. With the increase of the number of nodes, the complexity of the system increases exponentially, which makes it difficult to improve the performance of the current distributed quantum sensing system. Although there is no mature AI-enabled distributed quantum sensing scheme, it has become an inevitable trend to develop AI-based aided design and decision-making methods in the face of the optimization needs of this complex system.

## 5. Summary and outlook

In recent years, remarkable progress has been made in the application of artificial intelligence in the fields of quantum communication and quantum sensing, which has injected new vitality into the development of these cutting-edge technologies. In terms of quantum communication, artificial intelligence technology has greatly improved the performance and security of quantum key distribution, quantum storage and quantum network by means of parameter optimization, real-time feedback control and attack detection. In the field of quantum sensing, the introduction of artificial intelligence has opened up a new path to achieve high-precision and high-sensitivity quantum measurement. Quantum sensing technology uses the quantum state change of microscopic particles to achieve ultra-high sensitivity detection of time, magnetic field and other physical quantities, and its accuracy can be improved to atomic scale. Artificial intelligence enables quantum sensing, which can not only optimize the performance of sensors, but also further improve the measurement accuracy through data analysis.

Although the application of artificial intelligence in quantum communication has made a lot of progress, it still faces some challenges, such as the interpretability of algorithms. Some artificial intelligence algorithms are considered to be "black box" models, and their decision-making process is difficult to explain. This lack of interpretability may affect the reliability and security of quantum communication systems; Due to the difficulty of data acquisition and annotation, it is not easy to acquire a large amount of data for algorithm training in the actual system. In addition, most of the current quantum communication and quantum sensing systems rely on classical artificial intelligence algorithms, and future research can further explore the application of quantum artificial intelligence algorithms in quantum communication and quantum sensing. Quantum artificial intelligence algorithms are expected to break through the bottleneck of classical algorithms and achieve more efficient and secure quantum communication and quantum sensing systems by using the superposition and entanglement characteristics of quantum computing.

The miniaturization of quantum communication and quantum sensing systems, the integration of chips and multi-scenario applications such as unmanned aerial vehicles are also hot research directions, but they still face many challenges. The miniaturization and chipping of the system need to solve the problems of high cost and low integration of quantum chips, and at the same time ensure the high performance and stability of the system. The computing resources and storage capacity of lightweight chips may be limited, while the actual quantum communication and quantum sensing systems usually need to process a large amount of real-time data. How to achieve efficient algorithm deployment and system feedback control under the condition of limited computing power has become an urgent problem to be solved. At the same time, complex task environment, dynamic scenarios and multi-task coordination requirements make artificial intelligence aided design and decision-making an inevitable trend.

In addition, the integration of quantum communication and quantum sensing systems is an important direction for future technology development. Quantum communication provides absolutely secure information transmission capability, while quantum sensing shows great potential in high-precision measurement. The combination of the two, namely "Integrated sensing and communication", can realize the deep integration of quantum communication and quantum sensing, and provide more efficient and reliable solutions for tasks in complex environments. For example, quantum sensing technology can be used for high-precision environmental monitoring and positioning navigation, while quantum communication ensures the secure transmission of these data. Future research needs to further explore the deep integration of artificial intelligence with quantum communication and quantum sensing, develop more efficient, secure and practical quantum communication and quantum sensing systems, and lay a solid foundation for wide applications.